%% file: rg.tex
\documentstyle[twoside,fleqn,espcrc2,epsf]{article}

\newcommand{\me}[3]{{\langle#1\vert#2\vert#3\rangle}}

\newcommand{\vev}[1]{{\langle#1\rangle}}

\def\rsim{\mathrel{\raise2pt\hbox to 8pt{\raise -5pt\hbox{$\sim$}\hss{$>$}}}}

\def\CA{ {\cal A}}
\def\CC{ {\cal C}}
\def\CL{ {\cal L}}
\def\CP{ {\cal P}}
\def\CS{ {\cal S}}
\def\CU{ {\cal U}}

\newcommand{\NPB}[1]{{\sl Nucl. Phys.} {\bf B#1}}

\newcommand{\PRD}[1]{{\sl Phys. Rev. } {\bf D#1}}
\newcommand{\PLB}[1]{{\sl Phys. Lett.} {\bf #1B}}

\newcommand{\AmS}{{\protect\the\textfont2
  A\kern-.1667em\lower.5ex\hbox{M}\kern-.125emS}}

\title{Meson Form-factors and Wave-functions with Wilson fermions}

\author{Rajan Gupta, David Daniel and Jeffrey Grandy
        \address{T-8, MS B-285, Los Alamos National Laboratory,
        Los Alamos, New Mexico 87545, U.~S.~A.}
       }

\begin{document}

\begin{abstract}
Results for semi-leptonic form-factors for processes like $D \to K l \nu$
and  the Bethe-Salpeter amplitudes  (BSA) for pion and rho  mesons are
presented.    The form-factor  data    is  consistent   with  previous
calculations.  We find that the long distance fall-off of BSA for both
$\pi$ and $\rho$ is very well  fit by an exponential, but surprisingly
the effective mass governing this fall-off is lighter than the pion's.
Lastly,    by  studying  the  dependence  of    $\rho$ polarization on
separation direction we show that there  is  a measureable $l=2$ state
in addition to $l=0$ in the BSA for the rho.
\end{abstract}

\maketitle

\everypar{\looseness=-5}
\section{$D \rightarrow Kl\nu$ SEMI-LEPTONIC FORM-FACTOR}

Form-factors for semi-leptonic  decays  of heavy-light pseudoscalars
are expected to provide possibly the most stringent constraints on
CKM angles.  Over the last few years two groups have presented
results for $D \rightarrow Kl\nu$ and $D \rightarrow K^* l\nu$ decays
\cite{Bernardmff} \cite{Romamff}.
These   results  have  large  statistical  errors,   and,  in  certain
instances,  are in conflict.   To  resolve these discrepancies  and to
check for systematic errors, we have undertaken  further studies using
different methods. Here we report preliminary  results for the case $D
\rightarrow Kl\nu$.

All results presented in this talk were obtained using  35 lattices of
size $16^3   \times 40$ at   $\beta = 6.0$.   The  Wilson action quark
propagators were calculated on doubled lattices ($16^3  \times  40 \to
16^3 \times 80$) using Wuppertal  sources.  The quark  masses used are
$\kappa = 0.154$ and $0.155$, corresponding to pions of mass $700$ and
$560\     MeV$   respectively.      (For   further    details      see
Ref.~\cite{wilsonpheno}).  In    the   study   of  wavefunctions   the
convergence criteria (change per link) used for gauge fixing to either
Coulomb or Landau  gauge  is $  10^{-6}$.  The results are preliminary
and need to be confirmed on larger spatial lattices.

The matrix element of the vector current for a $D \rightarrow K$ transition
can be parameterized in terms of two form factors:
\begin{eqnarray}
\nonumber
\lefteqn{H_\mu \equiv
         \me{K^-(p_K)}{\bar s \gamma_\mu (1 - \gamma_5) c }{D^0(p_D)} } \\ &&
\label{mffdefeqn}
    \mbox{} \qquad \ =\  p_\mu f_+(q^2) + q_\mu f_-(q^2),
\end{eqnarray}
where $p=(p_D+p_K)$ and $q=(p_D-p_K)$ is the momentum  carried away by
the  leptons.  For the  vector current we  use three different lattice
transcriptions; the local current,   1-link  extended current and  the
conserved current.  The non-local currents  are symmetrized so that
they are defined at integer values of $t$.

\begin{table}[hbt]
\vspace{-0.3truein}
\caption{Raw lattice data for matrix elements $H_\mu$ with the 3 different
transcriptions of the vector current.}
\label{tab:MEmff}

\input t_me_ffP.tex

\vspace{-0.4truein}
\end{table}

The 3-point  correlator is  evaluated  as follows:   we  start  with a
Wuppertal  source   light  quark  propagator  and make   a  $\gamma_5$
insertion at zero 3-momentum at $t=32$.  We use the result as a source
for   a  second inversion   with the   charm    quark mass   fixed  at
$\kappa=0.135$.   This light-heavy  propagator  with  a   pseudoscalar
insertion at $t=32$  is then contracted with  a light quark propagator
at $t=1$ with a $\gamma_5$  to form the kaon.   The contraction at the
other end with the vector current is done  at all  intermediate times.
The initial  $D$  meson is therefore  always at rest  and  momentum is
inserted through the vector current.  We only consider the cases $\vec
p_K  = (0,0,0)$  and  $(0,0,1)$ even  though  Wuppertal  source  quark
propagators allow coupling to kaons of  all possible momenta.  This is
because the signal is poor for the higher momenta.

The raw lattice numbers for  the  three non-zero $H_\mu$ are  given in
Table 1.   There are two theoretical  issues that need to  be resolved
before one can extract $f_\pm$ from these numbers: the renormalization
constant for all three currents and the normalization of the heavy $c$
quark.  These effects could  be as large  as $20\%-30\%$ due to $O(a)$
corrections.  Ignoring  both these issues, $i.e.$  setting $Z_V = 1.0$
and not correction for the  heavy quark,  our results using  local and
conserved current are given in Table 2.  For momentum transfer $\vec p
= 0$ only $f_0$ is non-zero, while for $\vec p = (0,0,1)$  we get both
$f_\pm$.   These three  results are shown  in columns 2-4.  Within the
uncertainty of the statistical errors  ($20\% - 40\%$) the two results
are consistent and roughly agree with previous estimates.  Clearly, to
make progress  it is  important to  improve the  statistics and use  a
bigger  lattice,   and   also to reduce   $O(a)$    artifacts   in the
normalization of the vector current and of heavy quarks.

\begin{table}[hbt]
\vspace{-0.2truein}
\caption{Form-factor data with local (upper half) and conserved vector
current.  Errors are $\sim 20-40\%$.}
\label{tab:mff}
\input t_ffP.tex
\vspace{-0.45truein}
\end{table}

\vspace{-0.15 truein}
\section{BETHE-SALPETER AMPLITUDES}

The equal-time Bethe-Salpeter amplitude for the pion is defined as
\begin{equation}
\label{BSpidef}
\CA_\pi(\vec x) \ = \ \vev{0| \bar d(\vec x) \gamma_5 \CU(\vec 0, \vec x)
                      u(\vec 0) | \pi(\vec p, )}
\end{equation}
where $\CU(\vec 0, \vec x)$ is a path-ordered product of gauge links that
joins points $\vec x $ and $\vec 0$ and makes the amplitude gauge invariant.
This amplitude is given by the following ratio of 2-point correlators:
\begin{equation}
\label{pitwopoint}
%% \CA_\pi (\vec x, t)\ = \
 { { \vev{0| \bar d(\vec x;t) \gamma_5 \CU(\vec 0, \vec x;t) u(\vec 0;t)
  \bar u(\vec y; 0) \gamma_5 d(\vec y; 0) | 0} } \over
  { \vev{0| \bar d(\vec 0;t) \gamma_5 u(\vec 0;t)
  \bar u(\vec y; 0) \gamma_5  d(\vec y; 0) | 0} } } \ .
\end{equation}
There are two other related amplitudes that we consider; Coulomb gauge
($\CC_\pi$)  and  Landau  gauge  ($\CL_\pi$).   These are  obtained by
transforming the quark propagators to Coulomb (Landau) gauge, and then
calculating the  ratio given in Eq.~\ref{pitwopoint}\  with $\CU = 1$,
$i.e.$ without including the links.

Chu $et\ al.$  \cite{chunegele}\ investigated the  simplest version of
$\CU(\vec 0,\vec x)$,  $i.e.$  the straight  line path  between points
that lie along one of the lattice axis.  Previous calculations show
that $\vev{r^2}_\pi$ measured from the simplest gauge invariant BSA is
smaller than that obtained in either Coulomb or  Landau gauge and that
these are $0.3 -  0.5$ of the  charge radius  measured  in experiments
\cite{chunegele}\ \cite{hecht}. (A better probe is density-density
correlations as discussed in Ref.~\cite{chunegele}).
In this study we generalize $\CU$ to ``fat'' paths  made up of smeared
links and show that the resultant gauge invariant  BSA is broader than
fixed gauge  amplitudes.
% we suggest other ansatz for $\CU$.

\subsection{Gauge Invariant BSA with Smearing}

We use the APE  smearing method that was  first introduced to  enhance
the signal in glueball calculations \cite{APEsmearing}. In this method
each link in the spatial direction $i$ is replaced by the sum
\begin{equation}
\label{apesmearing}
U^{(1)}_i(\vec x, \vec x + \hat i)  =
\CP \bigg( U_i(\vec x, \vec x + \hat i)  +
           \sum_{i=1}^4  \CS_i(\vec x, \vec x + \hat i)  \bigg)
\end{equation}
where $\CS_i$ are the four spatial staples  shared by the  link $U_i$,
and the symbol $\CP$ implies that the sum is projected back  on to the
group SU(3).
% This  smearing step produces a  new lattice  of the same size  but
% made up   of  links which represent  locally  averaged gauge fields.
One can iterate  this  smearing     step as many   times as
necessary, using  the effective fields at   any step to  produce still
``fatter'' fields;  for example in the second  smearing step the right
hand side of Eq.~\ref{apesmearing}\ is  constructed from smeared links
produced in step one.  We specify the  smearing level by a superscript
on $\CA$, which will be $0-6$ corresponding  to  the original links and six
levels of smearing. We did not consider it appropriate  to go beyond 6
levels of smearing on a  lattice  of size  $16$ with periodic boundary
conditions.

\begin{figure}[t]
\vspace{-0.2truein}
\epsfysize=2.9in
\epsfxsize=2.5in
\epsfbox{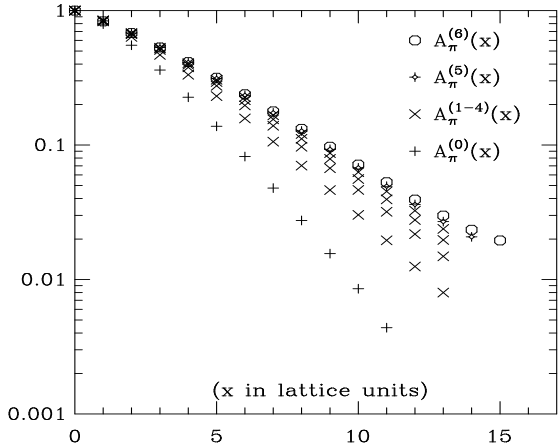}
\vspace{-0.15truein}
\caption{BSA for the pion at levels 0-6 of smearing.  Errors have been
suppressed for clarity}
\vspace{-0.2truein}
\label{fig:pionGI}
\end{figure}

There is  no unique answer  for  the  gauge-invariant BSA as different
choices  of  $\CU$ yield different results.   In Fig.   1 we show that
$\CA^{(i)}_\pi (\vec x, t=15)$ falls off less  rapidly as the smearing
level  $i$  is increased.   The  statistical errors  are  similar  for
smearing levels $1-6$, and the data show a rough convergence with $i$.
We therefore  use  results for $i=6$ as our  present best estimate for
the  gauge-invariant  BS amplitude.  The  second noteworthy feature is
that the large   $x$  behavior is   fit  well by an exponential    for
$\CA^{(1-6)}_\pi$ for $x  \geq 6$, while such a  behavior is  hard  to
extract from  $\CA^{(0)}_\pi$.   We   find that  $\CA^{(6)}_\pi   \sim
e^{-0.3x}$, a slower  fall-off  than  one  would  expect  as  $m_\pi =
0.365(6)$.  Qualitatively, the shape of  the BSA does  not change with
$t$,  though quantitatively it gets   significantly broader with  $t$,
reaching a steady state by about $t=15$.  This $t$  is somewhat larger
than $t = 10$ by which the correlator  is dominated by the lightest
state as shown in Ref.~\cite{wilsonpheno}\ using the same set of lattices.

We have also measured these amplitudes for non-zero momentum  with the
separation $\vec x$ taken  to  be  parallel  or perpendicular  to  the
direction of $\vec  p$.  This allowed  us to qualitatively verify that
the data show the expected Lorentz contraction.

\begin{figure}[t]
\vspace{-0.2truein}
\epsfysize=2.9in
\epsfxsize=2.6in
\epsfbox{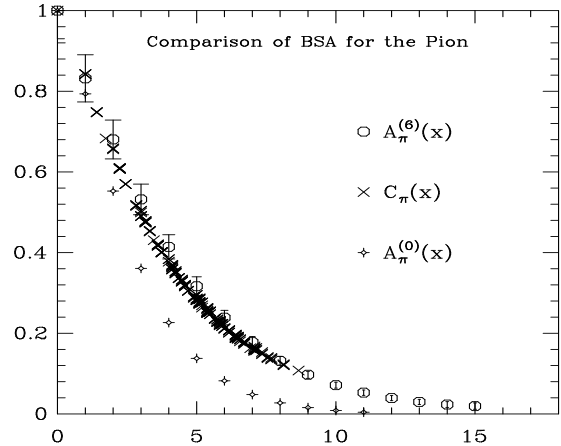}
\vspace{-0.15truein}
\caption{Comparison of the gauge-invariant BSA at smearing levels 0 and 6
with the Coulomb gauge result. We only show $\CC_\pi$ for $x_i \leq 6$.}
\vspace{-0.2truein}
\label{fig:compGIcg}
\end{figure}

\subsection{Comparison between gauge invariant, Landau, and Coulomb gauge BSA}

The  Coulomb and Landau gauge  amplitudes   have been measured for the
following relative separations: the anti-quark's position is varied in
a cube of size $9 \times 9 \times  9$ with respect  to the position of
the quark.  For each of these relative  separations we sum the quark's
position over  the time slice to  produce  a zero-momentum state.  The
data show   that  for  $x  >  6$ along  any  of the  axis  there  is a
significant  contamination from  wrap-around  effects due to  periodic
boundary   conditions.  These  effects can  be  taken into account  by
incorporating the contributions of all the mirror points.

The data show that $\CA^{(6)}$ is slightly broader than $\CL$ which in
turn is slightly broader than $\CC$.  On the other hand we find $\CL \rsim
\CC > \CA^{(0)}$   consistent with the  earlier results of Ref.~\cite{hecht}.
These two  features are illustrated  in Fig.  2 by data  for $\CC_\pi$,
$\CA^{(0)}_\pi$ and  $\CA^{(6)}_\pi$.  Thus $\CA^{(6)}$ is a  better probe of
quark/anti-quark distribution than Coulomb or  Landau gauge BSA.

%% In principle a
%% still better definition of $\CU$ is one  in which $\CU$ mimics a quark
%% propagator with the  same mass as a $u$  or  $d$. (This is easiest  to
%% visualize in the hopping parameter expansion where the propagator from
%% $\vec 0$ to $\vec x$ is a sum over paths).

\begin{figure}[t]
\vspace{-0.2truein}
\epsfysize=2.9in
\epsfxsize=2.6in
\epsfbox{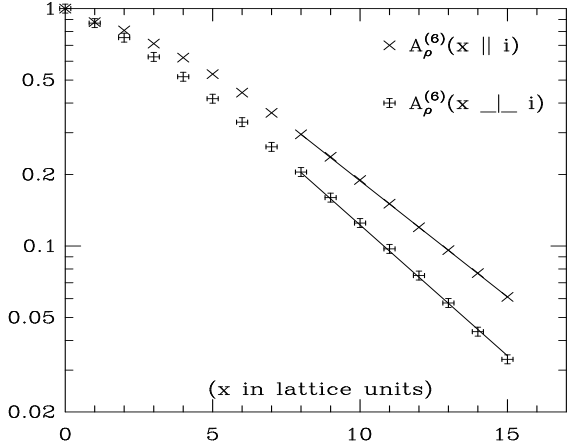}
\vspace{-0.15truein}
\caption{The BSA for the rho at $\kappa=0.154$ with polarization
axis $i$ $\|$ and $\bot$ to separation $\vec x$.}
\label{fig:rhopp}
\vspace{-0.2truein}
\end{figure}

\begin{figure}[t]
\vspace{-0.2truein}
\epsfysize=2.9in
\epsfxsize=2.65in
\epsfbox{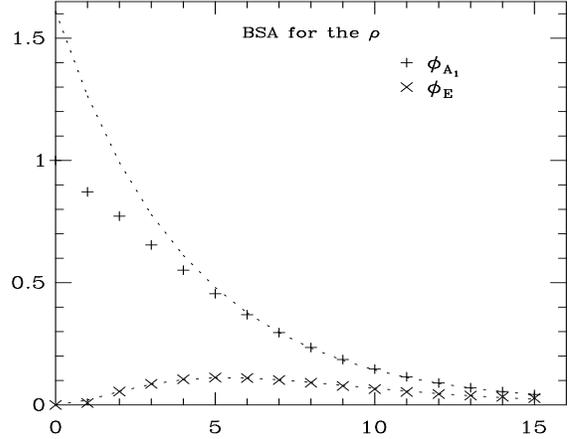}
\vspace{-0.15truein}
\caption{Results for $\phi_{A_1}(x)$ and $\phi_{E}(x)$ using the data
shown in Fig. 3.  The fits are described in Eq.~(6).}
\label{fig:rhostates}
\vspace{-0.2truein}
\end{figure}

\subsection{Polarizaton dependence of the $\rho$ BSA}

The $\rho$  meson wavefunction  is a linear  combination of  $l=0$ and
$l=2$ orbital angular momentum states.  On  the lattice the BSA
can be decomposed under the cubic group as
\begin{eqnarray}
\nonumber
\lefteqn{ \vev{0| \bar d(\vec x, t) \gamma_i \CU(\vec 0, \vec x;t) u(\vec 0,
t)
          | \rho(\vec 0, j)} } \\
\nonumber
  &=& {m_\rho^2 \over f_\rho}
    \bigg[
      \delta_{ij} \phi_{A_1}(\vec x)
      + \delta_{ij} \left( {x_i^2 \over \vec{x}\,^2} - {1\over3} \right)
        \phi_{E}(\vec x)  \\
\label{rhoWFlattice}
 & & \qquad  + (1-\delta_{ij}) {x_i x_j \over \vec{x}\,^2 } \phi_{T_2}(\vec x)
\bigg].
\end{eqnarray}
The functions  $\phi_{A_1}$, $\phi_{E}$ and  $\phi_{T_2}$  are scalars
under the cubic group with $E$ ($T_2$) labelling the 2 (3) dimensional
decomposition of the $l=2$ state.  Thus  lattice calculations allow us
to investigate, as a function of the  quark mass, the relative mixture
of $l=0$ and $l=2$ states, and  the restoration of rotational symmetry
by studying the three cases: (A) $i=j$ and $\vec x $ along $i$ ($\|$),
(B) $i=j$ and $\vec x $ perpendicular to $i$ ($\bot$), and (C) $i
\neq j$.

At present we  have only measured  the  BSA for $i=j$, and the results
for cases (A) and (B) at $\kappa = 0.154$ are shown  in  Fig. 3.  The
data show that for large separation the fall-off is extremely well fit
by  an exponential in  both  cases, with  a  rate of fall-off given by
$m_{\|} = 0.226(8)$ and $m_\bot = 0.263(7)$ respectively. Again, it is
interesting to note that the large $x$ fall-off is governed by a mass
that is lighter  than $m_\pi =  0.365$.

From the data shown in Fig. 3 we extract $\phi_{A_1}(x)$ and $\phi_{E}(x)$.
The results, shown in Fig. 4,  are fit to simple
hydrogen-like radial wavefunctions
\begin{eqnarray}
\nonumber
\phi_{A_1}(x) \ &=& \  1.61 \ e^{-0.242 r}      \\
\label{radialfits}
\phi_{E}(x)   \ &=& \  0.029 \ r^2 \ e^{-0.375 r} \ .
\end{eqnarray}
These functions give a good fit to
$\phi_{A_1}(x)$ for $r\geq6$ and $\phi_{E}(x)$ for $r\geq2$.
The results  at $\kappa =  0.155$ are qualitatively similar.  The data
for both $\phi_{A_1}(x)$  and $\phi_{E}(x)$ are  slighly broader,
though the difference is smaller than the statistical errors.

\medskip
\noindent{ACKNOWLEDGEMENTS:} These calculations were done on Cray YMP
using  time provided  by  LANL,  NERSC,   PSC and SDSC   Supercomputer
centers.  We thank DOE, NSF and Cray Research for their support.

\end{document}

%% file: t_me_ffP.tex
%% jk_mff in.mffP
%% /rg/rajan/=src/fits/dfit/LABELS.mffP
%% Reading output file
%% OUTPUT-fit_mff-in.mffP
%% Creator: rajan@qcd:fit_mff on Fri Sep 11 22:42:55 1992
%% Creator: rajan@qcd:fit_mff on Fri Sep 11 23:39:10 1992
%% \magnification=\magstep4
%% \hsize 7 truein
\def\CME{\mathord{\cal M} {\cal E}}

%
% header
%
% $$
\vbox{\hbox{\indent\vbox{\tabskip=0pt\offinterlineskip
\halign {\strut#& \vrule#\tabskip=4pt&
\hfil$#$\hfil&\vrule#&
\hfil$#$\hfil&\vrule#&
\hfil$#$\hfil&\vrule#
\tabskip=0pt\cr\noalign{\hrule}
%
% column labels
%
\omit&height2pt& && && & \cr
&&
&& \hbox{ \vtop{\hbox{$\kappa_1=0.154$}\smallskip\hbox{$\kappa_2=0.135$}}
}
&& \hbox{ \vtop{\hbox{$\kappa_1=0.155$}\smallskip\hbox{$\kappa_2=0.135$}}
}
& \cr
\omit&height2pt& && && & \cr
\noalign{\hrule}
%
% table
%
\omit&height1pt& && && & \cr
\omit&height1pt& && && & \cr
&& \hbox{ $\CME^{\rm loc.}(V_4,\vec p=0)$} && 1.18(11) && 1.19(16)   &\cr
\omit&height1pt& && && & \cr
% uncomment the next line for a rule separating rows
%\omit&height1pt& && && & \cr\noalign{\hrule}\omit&height1pt& && && & \cr
\omit&height1pt& && && & \cr
&& \hbox{ $\CME^{\rm loc.}(V_i,\vec p=1)$} && 0.51(11) && 0.72(25)   &\cr
\omit&height1pt& && && & \cr
% uncomment the next line for a rule separating rows
%\omit&height1pt& && && & \cr\noalign{\hrule}\omit&height1pt& && && & \cr
\omit&height1pt& && && & \cr
&& \hbox{ $\CME^{\rm loc.}(V_4,\vec p=1)$} && 0.83(30) && 0.84(40)   &\cr
\omit&height1pt& && && & \cr
% uncomment the next line for a rule separating rows
\omit&height1pt& && && & \cr\noalign{\hrule}\omit&height1pt& && && & \cr
\omit&height1pt& && && & \cr
&& \hbox{ $\CME^{\rm ext.}(V_4,\vec p=0)$} && 1.06(10) && 1.07(15)   &\cr
\omit&height1pt& && && & \cr
% uncomment the next line for a rule separating rows
%\omit&height1pt& && && & \cr\noalign{\hrule}\omit&height1pt& && && & \cr
\omit&height1pt& && && & \cr
&& \hbox{ $\CME^{\rm ext.}(V_i,\vec p=1)$} && 0.40(08) && 0.55(19)   &\cr
\omit&height1pt& && && & \cr
% uncomment the next line for a rule separating rows
%\omit&height1pt& && && & \cr\noalign{\hrule}\omit&height1pt& && && & \cr
\omit&height1pt& && && & \cr
&& \hbox{ $\CME^{\rm ext.}(V_4,\vec p=1)$} && 0.75(27) && 0.76(36)   &\cr
\omit&height1pt& && && & \cr
% uncomment the next line for a rule separating rows
\omit&height1pt& && && & \cr\noalign{\hrule}\omit&height1pt& && && & \cr
\omit&height1pt& && && & \cr
&& \hbox{ $\CME^{\rm con.}(V_4,\vec p=0)$} && 1.08(10) && 1.08(15)   &\cr
\omit&height1pt& && && & \cr
% uncomment the next line for a rule separating rows
%\omit&height1pt& && && & \cr\noalign{\hrule}\omit&height1pt& && && & \cr
\omit&height1pt& && && & \cr
&& \hbox{ $\CME^{\rm con.}(V_i,\vec p=1)$} && 0.35(08) && 0.46(16)   &\cr
\omit&height1pt& && && & \cr
% uncomment the next line for a rule separating rows
%\omit&height1pt& && && & \cr\noalign{\hrule}\omit&height1pt& && && & \cr
\omit&height1pt& && && & \cr
&& \hbox{ $\CME^{\rm con.}(V_4,\vec p=1)$} && 0.85(29) && 0.87(39)   &\cr
\omit&height1pt& && && & \cr
% uncomment the next line for a rule separating rows
%\omit&height1pt& && && & \cr\noalign{\hrule}\omit&height1pt& && && & \cr
\noalign{\hrule}
%
% close table
%
\cr}}}}
%$$

%% file: t_ffP.tex
%
% header
%
% \magnification=\magstep5
% \nopagenumbers
$$
\vbox{\hbox{\indent\vbox{\tabskip=0pt\offinterlineskip
\halign {\strut#& \vrule#\tabskip=4pt&
\hfil$#$\hfil&\vrule#&
\hfil$#$\hfil&\vrule#&
\hfil$#$\hfil&\vrule#&
\hfil$#$\hfil&\vrule#
\tabskip=0pt\cr\noalign{\hrule}
%
% column labels
%
\omit&height1pt&    &&           &&          &&           & \cr
&&                  &&f_0(\vec p = 0) &&f_+(\vec p = 1) &&f_-(\vec p = 1) &
\cr
\omit&height1pt&    &&           &&          &&           & \cr
\noalign{\hrule}
%
% table
%
\omit&height1pt&    &&           &&          &&           & \cr
\omit&height1pt&    &&           &&          &&           & \cr
&& \kappa_1=0.154   &&   0.98    &&   0.73   &&   -0.57   & \cr
\omit&height1pt&    &&           &&          &&           & \cr
% uncomment the next line for a rule separating rows
%\omit&height1pt& && && && & \cr\noalign{\hrule}\omit&height1pt& && && && &
% \cr
%
\omit&height1pt&    &&           &&          &&           & \cr
&& \kappa_1=0.155   &&   1.07    &&   0.85   &&   -0.98   & \cr
\omit&height1pt&    &&           &&          &&           & \cr
% uncomment the next line for a rule separating rows
\omit&height1pt& && && && & \cr\noalign{\hrule}\omit&height1pt& && && && & \cr
\noalign{\hrule}
\omit&height1pt&    &&           &&          &&           & \cr
\omit&height1pt&    &&           &&          &&           & \cr
&& \kappa_1=0.154   &&   0.90    &&   0.67   &&   -0.22   & \cr
\omit&height1pt&    &&           &&          &&           & \cr
% uncomment the next line for a rule separating rows
%\omit&height1pt& && && && & \cr\noalign{\hrule}\omit&height1pt& && && && &
% \cr
%
\omit&height1pt&    &&           &&          &&           & \cr
&& \kappa_1=0.155   &&   0.97    &&   0.76   &&   -0.41   & \cr
\omit&height1pt&    &&           &&          &&           &
\cr\noalign{\hrule}
% uncomment the next line for a rule separating rows
%\omit&height1pt& && && && & \cr\noalign{\hrule}\omit&height1pt& && && && &
% \cr
%
% close table
%
\cr}}}}
$$